%% file: main.tex
\documentclass[
	a4paper, 
	10pt, 
	unnumberedsections, 
	twoside, 
]{LTJournalArticle}

\usepackage[acronym]{glossaries}
\input{acronyms}
\makeglossaries
\usepackage{biblatex}
\runninghead{} 

\footertext{\textit{RISC-V Summit Europe, Paris, 12-15th May 2025}} 

\title{Towards a Base-Station-on-Chip: RISC-V Hardware Acceleration for wireless communication.} 

\author{%
	Javier Acevedo\textsuperscript{1} and Frank H. P. Fitzek\textsuperscript{1,2} \thanks{Corresponding author: \href{mailto:jane@smith.com}{\tt javier.acevedo@tu-dresden.de}. \\
    \textbf{Acknowledgment:} This research has been partially funded by the Federal
Ministry of Education and Research (BMBF) under grant 01IS17044 High-Tech Strategy 2025 (HTS2025), as part of the Software Campus project “RISC-ARA”.
}
}

\date{\footnotesize\textsuperscript{\textbf{1}}Deutsche Telekom Chair of Communication Networks, TU Dresden\\ \textsuperscript{\textbf{2}}Centre for Tactile Internet with Human-in-the-Loop (CeTI)}

\renewcommand{\maketitlehookd}{%
	\begin{abstract}
		\noindent The evolution of 5G and the emergence of 6G wireless communication systems impose higher demands for computing capabilities and lower power consumption in the front-end and processing circuitry. Furthermore, the incorporation of \gls{ai}/\gls{ml} in the \gls{ran} introduces heightened computational needs and stringent low-latency requirements for both training and inference. The concept of a \gls{bsoc} addresses those demands by consolidating of the signal processing, neural network computations and network management functions into a single chip. This new computing platform relies on a sophisticated  hardware/software co-design to optimize performance, power efficiency, and scalability, enabling a compact, yet adaptable and intelligent base station solution for next-generation wireless networks. This research investigates the efficient implementation of conventional \gls{ce}, \gls{mmimo}, and beamforming kernels on a state-of-the-art RISC-V vector \gls{dsp} to capitalize on \gls{dlp}. Moreover, it explores how \gls{rvv} combined with custom instructions can effectively address the throughput and latency demands of LOW \gls{phy} kernels.
	\end{abstract}
}

\begin{document}

\maketitle 
\section{Introduction}

\begin{figure*}[hbt!]
    \centering
    \includegraphics[width=0.95\textwidth]{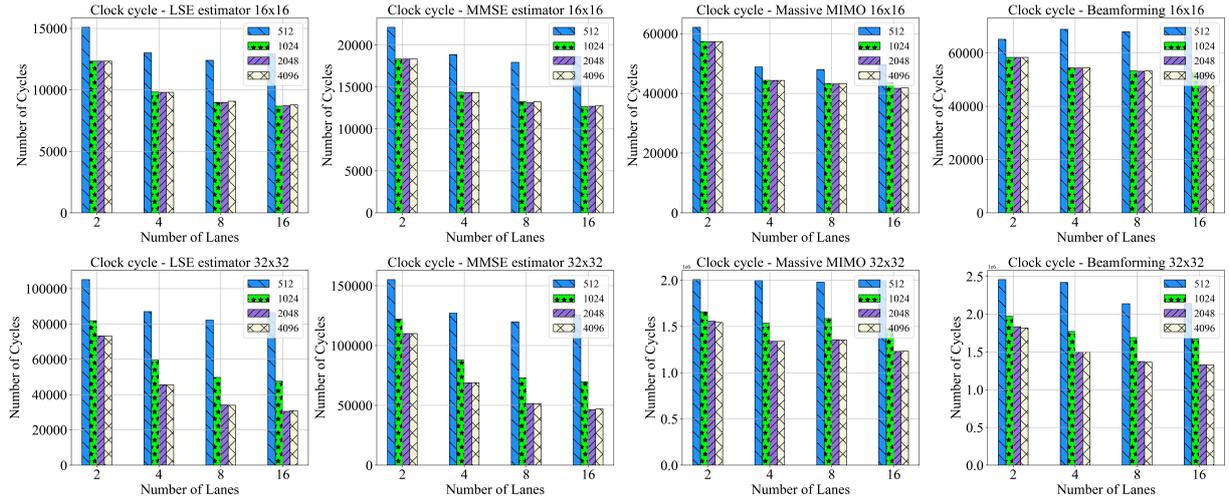}
    \caption{Clock cycle counts for the execution of the \gls{mmse} \gls{ce}, \gls{lse} \gls{ce}, \gls{mmimo} and beamforming algorithms employing
different matrix sizes ($16 \times 16$ and $32 \times 32$) and $VLEN$ values (512, 1024, 2048, and 4096 bit). }
    \label{fig:radio} 
\end{figure*}

The advent of the \gls{openran} has transformed profoundly wireless cellular networks by decoupling hardware and software into modular components, which are interconnected via open interfaces like \gls{ecpri}. This disaggregation fosters innovation by enabling the virtualization of the \gls{ran} function on \gls{cots} servers, reducing reliance on proprietary vendor equipment. Nevertheless, \gls{cots} hardware is built on fixed-length \gls{simd} architectures, which cannot be adapted to fulfill the computational demands of the LOW \gls{phy}  signal processing algorithms. In contrast, the open-source RISC-V Instruction Set Architecture (ISA) supports customizable vector lengths, facilitating the design of specialized vector processors tailored for wireless communication tasks such as \gls{ce}, beamforming, and \gls{mmimo}. The aforementioned kernels represent the main signal processing algorithms performed at the \gls{oru} of a \gls{openran}-compliant base station. Therefore, congregating the execution of those kernels into a single chip promotes new computing platforms for small cells: the \gls{bsoc}.  This work aims to investigate how RISC-V can be leveraged to develop hardware accelerators that meet the stringent throughput, latency, and power requirements of next-generation \gls{6g} base stations.

LOW \gls{phy} processing in base stations involves computationally intensive operations, including massive matrix multiplications, matrix inversions, and \gls{fft}/iFFT computations, critical for real-time \gls{ce} and \gls{mmimo}. The computational complexity of these algorithms scales significantly with the number of antenna elements and system configuration. The RISC-V \gls{isa} flexibility allows for customized solutions to address this complexity efficiently. In this study, we utilize the state-of-the-art RISC-V-based Ara processor from the PULP group to implement hardware accelerators for LOW \gls{phy} algorithms. Our objectives are two-fold:
\begin{itemize}
    \item Assess speedup in kernel execution throughout data parallelization.
    \item Develop custom hardware modules optimized for each signal processing kernel.
\end{itemize}

\section{Approach}
The Ara processor is a high-performance, open-source, RISC-V core designed for parallel processing. LOW \gls{phy} are characterized for having multiple arithmetic operations, which can be vectorized and hence accelerated by distributing the computation over multiple parallel lanes. For \gls{ce}, we implemented \gls{lse} and \gls{mmse} kernel to determine the channel matrix $H$, by solving the equation $H = YX^{-1}$ throughout the calculation of the least squares and statistics. On the other hand, we employed the Cooley-Tukey to compute the radix 4 \gls{fft}.  In the case of \gls{mmimo}, we calculated the \gls{zf} precoder, $W$, by solving the equation given by $W = H^{\textbf{H}}(HH^{\textbf{H}})^{-1}$, where $\textbf{H}$ represents the hermitian or conjugate transpose of the channel matrix $H$. Additionally, we have done an extension for the digital beamforming to construct the channel matrix including the steering vectors. In such a manner, we could represent the phase and amplitude, which are required to be applied to each antenna element to direct the beams precisely.

\section{Preliminary results}
In this work, we provide a C-based software implementation of the aforementioned wireless communication kernels. We simulated and evaluated each kernel by adjusting the number of lanes and measuring the number of clock cycles required by the Ara core to perform the computation. This approach allows us to observe how vectorization impacts the execution speedup of each kernel, consistent with the findings presented in \cite{ce1, ce2}.

Our evaluation measures the clock cycle count of the \gls{lse} and \gls{mmse} \gls{ce}, \gls{mmimo}, and beamforming. Figure \ref{fig:radio} illustrates these results across various vector registers lengths, $VLEN$, and number of lanes. The Ara core supports 64-bit values, which dictates he number of elements processed in parallel within the hardware lanes. The bigger the $VLEN$, the higher the number of elements employed for computation within a clock cycle. Matrix sizes are varied to demonstrate their effect on the number of parallel operations and, consequently, the total clock cycles needed to complete each kernel's computation.

\section{Conclusion and Outlook}
In this study, we introduced the initial findings from the software implementation of multiple wireless communication kernels on a RISC-V vector processor. By leveraging the \gls{dlp} inherent in these algorithms, we achieve a reduction in the clock cycle count as the number of parallel processing lanes increases. Future work will include a custom hardware implementation of some of those kernels and their integration into the Ara processor via AXI interfaces \cite{sdr1}. Hence, the development of tailored instructions to provide support to that hardware is also planned \cite{simd1}.

\printbibliography 

\end{document}

%% file: acronyms.tex
\newacronym{5g}{5G}{5th Generation Cellular Networks}
\newacronym{6g}{6G}{6th Generation Cellular Networks}
\newacronym{5gacia}{5G-ACIA}{5G Alliance for Connected Industries and Automation}

\newacronym{ack}{ACK}{Acknowledgement Message}
\newacronym{ai}{AI}{Artificial Intelligence}
\newacronym{alu}{ALU}{Arithmetic Logic Unit}
\newacronym{ar}{AR}{Augmented Reality}
\newacronym{asic}{ASIC}{Application Specific Integrated Circuit}
\newacronym{awgn}{AWGN}{Additive White Gaussian Noise}

\newacronym{be}{BE}{Best Effort}
\newacronym{bg}{BG}{Background}
\newacronym{bsoc}{BSoC}{Base Station on Chip}
\newacronym{bw}{BW}{Bandwidth}
\newacronym{ce}{CE}{Channel Estimation}
\newacronym{cir}{CIR}{Channel Impulse Response}
\newacronym{cfr}{CFR}{Channel Frequency Response}
\newacronym{csi}{CSI}{Channel State Information}
\newacronym{csirs}{CSI-RS}{Channel State Information Reference Signal}
\newacronym{cots}{COTS}{Commercial-Off-The-Shelf}
\newacronym{ccdf}{CCDF}{Complementary \gls{cdf}}
\newacronym{cmac}{CMAC}{Complex Multiply Accumulate}
\newacronym{cdf}{CDF}{Cumulative Distribution Function}

\newacronym{dif}{DIF}{Decimation In Frequency}
\newacronym{dit}{DIT}{Decimation In Time}
\newacronym{dlp}{DLP}{Data Level Parallelism}
\newacronym{dmrs}{DM-RS}{Demodulation Reference Signal}
\newacronym{dut}{DuT}{Device under Test}
\newacronym{dft}{DFT}{Discrete Fourier Transform}
\newacronym{dpdk}{DPDK}{Data Plane Development Kit}
\newacronym{dsp}{DSP}{Digital Signal Processors}

\newacronym{ecpri}{eCPRI}{evolved Common Public Radio Interface}
\newacronym{fft}{FFT}{Fast Fourier Transform}
\newacronym{fpga}{FPGA}{Field Programmable Gate Array}
\newacronym{fpu}{FPU}{Floating Point Unit}

\newacronym{gps}{GPS}{Global Positioning System}
\newacronym{gcl}{GCL}{Gate Control List}
\newacronym{gfdm}{GFDM}{Generalized Frequency Division Multiplexing}
\newacronym{gptp}{gPTP}{Generic \gls{ptp}}
\newacronym{gpu}{GPU}{Graphic Processing Unit}
\newacronym{gm}{GM}{grandmaster}

\newacronym{ifg}{IFG}{Inter-Frame Gap}
\newacronym{isa}{ISA}{Instruction Set Architecture}
\newacronym{i40}{I4.0}{Industry 4.0}

\newacronym{kpi}{KPI}{Key Performance Indicator}
\newacronym{L2}{L2}{\gls{osi} Layer 2}
\newacronym{L3}{L3}{\gls{osi} Layer 3}
\newacronym{lse}{LSE}{Least-Squares}
\newacronym{mac}{MAC}{Multiply-Accumulate Operation}
\newacronym{ml}{ML}{Machine Learning}
\newacronym{mmimo}{mMIMO}{massive Multiple Input Multiple Output}
\newacronym{mmwave}{mmWave}{Millimeter Wave}
\newacronym{mmse}{MMSE}{Minimum Mean Square Error}
\newacronym{mse}{MSE}{Minimum Square Error}
\newacronym{mul}{MUL}{Multiplication Vector}

\newacronym{nic}{NIC}{Network Interface Card}
\newacronym{nr}{NR}{New Radio}

\newacronym{ofdm}{OFDM}{Orthogonal Frequency Division Multiplexing}
\newacronym{osi}{OSI}{Open Systems Interconnection}
\newacronym{openran}{Open RAN}{Open Radio Access Network}
\newacronym{oru}{O-RU}{Open Radio Radio Unit}

\newacronym{pcg}{PCG}{Program Counter Generation}
\newacronym{ptp}{PTP}{Precision Time Protocol}
\newacronym{p2p}{P2P}{Peer-to-Peer}
\newacronym{ptrs}{PT-RS}{Phase Tracking Reference Signal}
\newacronym{pdsch}{PDSCH}{Physical Downlink Shared Channel}
\newacronym{phy}{PHY}{Physical Layer}
\newacronym{prb}{PRB}{Physical Resource Block}
\newacronym{pcp}{PCP}{Priority Code Point}

\newacronym{qos}{QoS}{Quality of Service}

\newacronym{ran}{RAN}{Radio Access Network}
\newacronym{rb}{RB}{Resource Block}
\newacronym{re}{RE}{Resource Element}
\newacronym{rg}{RG}{Resource Grid}
\newacronym{risc}{RISC}{Reduced Instruction Set Computer}
\newacronym{rx}{RX}{receiver}
\newacronym{rfc}{RFC}{Request for Comment}
\newacronym{rtl}{RTL}{Register-Transfer Level}
\newacronym{rvv}{RVV}{RISC-V Vector Extensions}

\newacronym{scs}{SCS}{Sub-carrier Spacing}
\newacronym{simd}{SIMD}{Single Instruction Multiple Data}
\newacronym{siso}{SISO}{Single-Input Single-Output}
\newacronym{snr}{SNR}{Signal-Noise Ratio}
\newacronym{SRP}{SRP}{Stream Reservation Protocol}
\newacronym{SDU}{SDU}{Service Data Unit}
\newacronym{SLA}{SLA}{Service Level Aggreement}

\newacronym{tas}{TAS}{Time-Aware Shaper}
\newacronym{tsn}{TSN}{Time-Sensitive Networking}
\newacronym{tx}{TX}{transmitter}

\newacronym{ul}{UL}{Uplink}
\newacronym{urllc}{URLLC}{Ultra-Reliable Low-Latency Communication}
\newacronym{udp}{UDP}{User Datagram Protocol}

\newacronym{vrf}{VRF}{Vector Register File}
\newacronym{vr}{VR}{Virtual Reality}

\newacronym{zf}{ZF}{Zero-Forcing}